\documentclass{article}

\usepackage{times}
\usepackage{xcolor}
\usepackage{soul}
\usepackage[utf8]{inputenc}
\usepackage[small]{caption}
\usepackage{amsmath}
\usepackage{amssymb}
\usepackage{algorithm}
\usepackage[noend]{algpseudocode}	
\usepackage{graphicx}
\usepackage{tikz}
\usetikzlibrary{arrows,automata}
\usepackage{subcaption}
\usepackage{tabularx}
\usepackage{multirow}
\usepackage{multicol}
\usepackage{booktabs}
\usepackage{url}
\usepackage{enumitem}
\usepackage{natbib}
\newtheorem{theorem}{Theorem}
\newtheorem{example}{Example}

\newcommand{\nop}[1]{}

\def\calG{\mathcal{G}}

\def\calR{\mathcal{R}}
\def\calO{\mathcal{O}}
\def\calW{\mathcal{W}}

\def\id{\mathit{id}}
\def\scc{\mathit{scc}}
\def\status{\mathit{status}}

\def\np{\mathsf{NP}}
\def\Ptime{\mathsf{P}}
\def\coP{{co}\mbox{-}\mathsf{P}}
\def\ptime{\mathsf{P}}

\def\nl{\mathsf{NL}}
\def\conl{\mathsf{co}\mbox{-}\mathsf{NL}}

\newcommand{\problem}[3]{
\begin{center}
{\small 
\begin{tabularx}{0.98\columnwidth}{ll}
\toprule
\multicolumn{2}{c}{\textsc{#1}} \\
\midrule
{\bf Instance:}   & \parbox[t]{0.7\columnwidth}{#2\vspace*{1mm}}  \\
{\bf Question:}& \parbox[t]{0.7\columnwidth}{#3\vspace*{.5mm}} \\ 
\bottomrule
\end{tabularx}
}
\end{center}
}

\title{Computing the Schulze Method for Large-Scale Preference Data Sets}
\date{}
\author{
Theresa Csar \and Martin Lackner \and Reinhard Pichler \\
TU Wien, Austria\\
\{csar, lackner, pichler\}@dbai.tuwien.ac.at
}

\begin{document}

\maketitle

\begin{abstract}
The Schulze method is a voting rule widely used in practice and enjoys many positive axiomatic properties.
While it is computable in polynomial time, its straight-forward implementation does not scale well for large elections.
In this paper, we develop a highly optimised algorithm for computing the Schulze method with Pregel, a framework for massively parallel computation of graph problems, and demonstrate its applicability for large preference data sets.
In addition, our theoretic analysis shows that the Schulze method is indeed particularly well-suited for parallel computation, in stark contrast to the related ranked pairs method. More precisely we show that winner determination subject to the Schulze method is NL-complete, whereas this problem is P-complete for the ranked pairs method.

\textbf{This is an updated version of the original 2018 IJCAI conference publication. It corrects the P-completeness proof for the ranked pairs method.}
\end{abstract}

\section{Introduction}

Preference aggregation is a core problem of computational social choice: how to aggregate potentially conflicting preferences of agents into a collective preference ranking or to identify  most preferred alternatives.
This problem occurs both in a technical context where agents correspond to machines or programs (e.g., multi-agent systems, group recommendation system, aggregation of information sources), and in a social context where people strive for a joint decision (e.g., political voting, group decision making).
The Schulze method~\citep{schulze2003new,Schulze} is a preference aggregation method---or \emph{voting rule}---with a particularly attractive set of properties\footnote{We note, however, that other voting rules satisfy similar or even stronger axiomatic properties, for example those defined by the minimal covering set, the uncovered set, or the bipartisan set~\citep{brandt2015handbook}.}. For example, the Schulze method satisfies the independence of clones criterion, i.e., if a candidate joins that is virtually indistinguishable from another candidate 
(i.e., a ``clone'' joins), then the relative ranking of alternatives in the output remains unchanged. Furthermore, the Schulze method is Condorcet-consistent and monotonic~\citep{Schulze}.
Partially due to these advantages, the Schulze method enjoys significant popularity and is widely used in group decisions of open-source projects and political groups\footnote{\url{https://en.wikipedia.org/wiki/Schulze_method#Users}}.

Preference aggregation is often a computationally challenging problem.
It is therefore fortunate that the Schulze method can be computed in polynomial time. More precisely, given preference rankings of $n$ voters over $m$ candidates, it requires $\calO(nm^2+m^3)$ time to compute the output ranking of all $m$ candidates.
However, while this runtime is clearly feasible for decision making in small- or medium sized groups, it does not scale well for a larger number of candidates due to the cubic exponent.
Instances with a huge number of alternatives occur in particular in technical settings, such as preference data collected by e-commerce applications or sensor systems.
If the number of alternatives is in the thousands, the classical algorithm for computing the Schulze method (based on the Floyd--Warshall algorithm) quickly reaches its limits.

The goal of our paper is to design a fast, parallel algorithm that enables preference aggregation via the Schulze method on large-scale data sets.
As a first step, we perform a worst-case complexity analysis and show that the Schulze Method is indeed suitable for parallel computation: we prove that the problem of computing a Schulze winner (i.e., the top-ranked alternative) is $\nl$-complete. We contrast this result by showing that the related ranked pairs method, which has similar axiomatic properties, is likely not to allow for effective parallel computation: 
here, we show that the winner determination problem is $\ptime$-complete.

Building on this theoretic foundation, we design a parallel algorithm for computing Schulze winners in the Pregel framework. Pregel~\citep{malewicz2010pregel} is a framework for cloud-based computation of graph problems. In Pregel, parallelization happens on the level of vertices, i.e., each vertex constitutes an independent computation unit that communicates with neighbouring vertices. 
The Schulze method is based on the weighted majority graph, i.e., it does not require actual rankings as input but rather pairwise majority margins for each pair of vertices.
Hence, computing the Schulze method is a problem very suitable for the Pregel framework.

As the main contribution of this paper, we present a highly optimised Pregel-based parallel algorithm for computing Schulze winners. This algorithm can easily be adapted to also computing the top-$k$ alternatives.
We demonstrate the effectiveness of our optimisations in an experimental evaluation.
We use daily music charts provided by the Spotify application to generate data sets with up to 18,400 alternatives; the corresponding weighted tournament graphs have up to 160 million weighted edges.
We show that such data sets can be computed in the matter of minutes and demonstrate that runtimes can be significantly reduced by an increase in parallelization.
Thus, our algorithm enables the application of the Schulze method in data-intensive settings.

\smallskip
\noindent
{\bf Structure and main results}.
We recall some basic notions and results in Section~\ref{sect:preliminaries}. 
A conclusion and 
discussion of future directions for research
are given in Section~\ref{sect:conclusion}. Our main results, which are detailed 
in Sections \ref{sect:complexity} -- \ref{sect:experiments}, are as follows: 
\\[1.1ex]
$\bullet$ \ In Section \ref{sect:complexity} we carry out a complexity-theoretic analysis of winner determination for the Schulze method and for the ranked pairs method. We thus establish a significant difference between the two methods in that the former is 
in $\nl$ whereas the latter is $\ptime$-hard. This explains why the Schulze method is very well-suited for parallelization.
\\[1.1ex]
$\bullet$ In Section \ref{sect:schulze}, we present a new parallel algorithm for the winner determination according to the Schulze method, using the vertex-centric algorithmic paradigm of Pregel. 
\\[1.1ex]
$\bullet$ In Section \ref{sect:experiments}, we report on experimental results with our Pregel-based algorithm. 
The empirical results confirm that parallelization via a vertex-centric approach indeed works very well in practice.


\smallskip

\noindent
{\bf Related work.}
We briefly review related work on algorithms for preference aggregation and winner determination. 
Most work in this direction is focused on $\np$-hard voting rules, in particular the 
Kemeny rule (see, e.g.,~\citep{conitzer2006improved,betzler2014theoretical,schalekamp2009rank,ali2012experiments}).

For some  voting rules the complexity changes depending on whether a fixed tie-breaking order is used. 
This is the case for the STV rule, where the winner determination problem is $\np$-hard if no tie-breaking order is specified~\citep{conitzer2009preference} (i.e., for the decision problem whether there is a tie-breaking order such that a distinguished candidate wins subject to this tie-breaking order), but STV is $\ptime$-complete for a fixed tie-breaking order~\citep{AAAI17}.
Similarly, the ranked pairs method is $\np$-hard to compute without specified tie-breaking~\citep{brill2012price}. 
We show in this paper that ranked pairs winner determination for a fixed tie-breaking order is $\ptime$-complete.
Recent work by~\citet{jiang2017practical} has considered the $\np$-hard variants of STV and ranked pairs and established fast algorithms for these problems.
The use of parallel algorithms for winner determination has been previously studied by~\citet{AAAI17} in the MapReduce framework, in particular for tournament solution concepts.
Finally, we remark that \citet{parkes2012complexity} considered
similarities and differences of the ranked pairs and Schulze method in the context of strategic voting.

\section{Preliminaries}
\label{sect:preliminaries}

A directed graph (digraph) is a pair $(V,E)$ with $E\subseteq V\times V$.
A path (from $x_1$ to $x_k$)  
is a sequence $\pi = (x_1, \dots, x_k)$ of vertices with 
$(x_i,x_{i+1}) \in E$ for every $i \in \{1, \dots, k-1\}$. We call $\pi$ 
a cycle, if $x_1 = x_k$. A digraph without cycles is referred to as DAG (directed acyclic graph).

A set $X \subseteq V$ of vertices is called strongly connected if for every pair $(a,b)$ of vertices in $X$, there is a path from $a$ to $b$ and from $b$ to $a$. 
If $X$ is maximal with this property, we call it a strongly connected component (SCC).

A weighted digraph is a triple $(V,E,w)$ with $(V,E)$ being a digraph and the function 
$w\colon E\to [0,\infty)$ 
assigning (non-negative) weights to edges. 

\subsection{Voting}
Let $A$ be a set of alternatives (or candidates) and $N=\{1,\dots,n\}$ a set of voters.
A preference profile $P=(\succeq_1,\dots,\succeq_n)$ contains the preferences of the voters. We require $\succeq_1$, $\dots$, $\succeq_n$ to be weak orders on $A$, i.e., transitive and complete relations. We write $\succ_i$ to denote the strict part of~$\succeq_i$.

We now define several concepts based on a given preference profile $P$.
We say that alternative $a$ dominates alternative $b$ if more voters prefer $a$ to $b$ than $b$ to $a$, i.e., $|i\in N: a\succ_i b|>|i\in N: b\succ_i a|$; let $D_P\subseteq A\times A$ denote the corresponding dominance relation.
The (strict) \emph{dominance graph} is the digraph $(A,D_P)$, i.e., there exists an edge from vertex $a$ to $b$ if and only if $a$ dominates $b$.
The \emph{majority margin} of two candidates $a,b$ is defined as $\mu_P(a,b)=|i\in N: a\succ_i b|-|i\in N: b\succ_i a|$.
Let $\calW_P= (A,E_P,\mu'_P)$ with $E_P=\{(a,b)\in A^2\colon \mu_P(a,b)>0\}$ and $\mu'_P$ being the restriction of $\mu_P$ to $E_P$.
Note that $\calW_P$ is a weighted digraph; we refer to it as the \emph{weighted tournament graph of $P$}.
We refer to elements of $A$ interchangeably as candidates or vertices.

The \emph{Schwartz set} is defined as 
the union of all non-dominated SCCs
in the dominance graph.
The \emph{Schulze method}~\citep{Schulze} is a refinement of the Schwartz method. 
Its definition depends on \emph{widest paths} in weighted graphs.
Let $(A,E,\mu)$ be a weighted tournament graph.
A path $(x_1,\dots,x_k)$  has 
\emph{width $\alpha$} if \[\min_{i\in\{1,\dots,k-1\}}\mu(x_i,x_{i+1})= \alpha.\]
A \emph{widest path from $a$ to $b$} is a path from $a$ to $b$ of maximum width; let $p(a,b)$ denote the width of such a path.
According to the Schulze method, an alternative $a$ beats 
alternative $b$ if there is a wider path from $a$ to $b$ than from $b$ to $a$, i.e., if $p(a,b)>p(b,a)$.
An alternative $a$ is a Schulze winner if there is no alternative $b$ that beats $a$.
The Schulze method can also be used to compute an output ranking, which is defined by the relation $(a,b)\in R$ if and only if $p(a,b)\geq p(b,a)$.
It can be shown that $R$ is a weak order~\citep{Schulze}.
The Schulze winners are exactly the top-ranked alternatives in $R$.

We define the \emph{ranked pairs method}~\citep{tideman1987independence} subject to a fixed tie-breaking order\footnote{We note that the ranked pairs method satisfies the independence of clones property only for specific tie-breaking orders~\citep{zavist1989complete}; our results are in particular applicable to these distinguished tie-breaking orders.} $T$, 
which is a linear order of the candidates.
The ranked pairs method creates a ranking, 
starting with an empty relation $R$.
All pairs of candidates are sorted according to their majority margin and ties are broken according to $T$. Then, pairs of candidates are added to the relation $R$ in the sorted order (starting with the largest majority margin). 
However, a pair is omitted if it would create a cycle in $R$.
The final relation $R$ is a ranking of all alternatives; the top-ranked alternative is the winner (subject~to~$T$). 

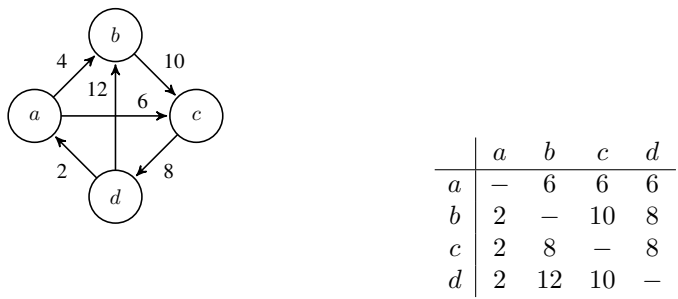
\begin{figure}
\centering
    \begin{subfigure}{0.5\columnwidth}
        \centering
        \begin{tikzpicture}[->,>=stealth',shorten >=1pt,auto,node distance=1.9cm,
                    semithick,scale=0.5, every node/.style={scale=0.8}]
  \tikzstyle{every state}=[inner sep=0.1cm, outer sep=0]

  \node[state] (A)                    {$a$};
  \node[state]         (B) [above right of=A] {$b$};
  \node[state]         (D) [below right of=A] {$d$};
  \node[state]         (C) [below right of=B] {$c$};
  \path (A) edge              node {4} (B)
            edge              node [pos=0.75] {6} (C)
        (B) edge              node {10} (C)
        (C) edge              node {8} (D)
        (D) edge              node {2} (A)
            edge              node [pos=0.75] {12} (B);
\end{tikzpicture}
    \end{subfigure}%
    ~\begin{subfigure}{0.45\columnwidth}
        \centering 
        $\begin{array}{c|cccc}
        & a & b & c & d \\ \hline
        a & - & 6 & 6 & 6 \\ 
        b & 2 & - & 10 & 8 \\ 
        c & 2 & 8 & -  & 8  \\ 
        d & 2 & 12 & 10 & -
        \end{array}$ 
    \end{subfigure}
    \caption{A weighted tournament graph and its widest paths.}\label{fig:tournament}
\end{figure}

\begin{example}
{\em 
Consider the weighted tournament graph displayed in Figure~\ref{fig:tournament}
on the left (originally by~\citep{schulze2003new}). 
The table in Figure~\ref{fig:tournament}
on the right shows the widest paths between any two vertices.
The unique {\it Schulze winner} is candidate $a$, having a path of width 6 to every other candidate, whereas all incoming paths to vertex $a$ have width $2$. 

The unique {\it ranked pairs winner} is $d$ (independently of the chosen tie-breaking order). It is obtained by inspecting the edges in descending order of weights and retaining an edge only if it does not cause a cycle. We thus retain 
$(d,b),(b,c)$, omit $(c,d)$, and retain $(a,c),(a,b), (d,a)$. 
Hence, $d$ is the only vertex without incoming edge in the resulting DAG.

The digraph in Figure~\ref{fig:tournament} is strongly connected. Hence, 
the set of {\it Schwartz winners} is the entire SCC $\{a,b,c,d\}$.
\hfill $\diamond$
}
\end{example}

\subsection{Cloud Computing Frameworks}
Cloud computing algorithms are based on the concept of splitting problem instances into small parts and performing computations on those parts using independent computation nodes. 
This approach gained popularity with the MapReduce framework \citep{DeanG08}, which is ideal for distributed batch processing; for a general overview of MapReduce and its variants we refer the reader to a survey by \citep{sakr2013family}. 
%
The related Pregel framework~\citep{malewicz2010pregel} was introduced 
specifically for big data problems on huge graphs.

Pregel algorithms are vertex-centric computations. They are often described as 'think-like-a-vertex' algorithms,
where each vertex acts as an independent computation entity. 
This means that the vertices are distributed among the nodes of the cluster and the computations at each vertex can be performed in parallel. The vertices exchange information by sending messages to each other along the edges of the graph. Moreover, each vertex can store its own local information. 

The Pregel computation works in supersteps;  
at the beginning of each superstep, the vertices read messages sent by other vertices in the previous superstep. 
If a vertex does not receive any messages, it is set inactive (but can be reactivated by messages in following supersteps); Pregel programs terminate if all vertices are inactive.
After receiving messages, a vertex performs its vertex program, in which 
the information stored at the vertex can be changed and messages can be sent to 
other vertices.
Finally, a message combiner can be used to collect and combine messages in order to optimise the communication between vertices.

\section{Computational Complexity}
\label{sect:complexity}
In this section, we classify the computational complexity of 
the Schulze method and the ranked pairs method. 
In both cases, we do not need a preference profile as input but only the corresponding weighted majority graph.
We thus define the winner determination problems for 
$\calR \in \{$Schulze, Ranked Pairs$\}$ as follows:

\problem{$\calR$ Winner Determination}{a weighted tournament graph $\calW=(A,E,\mu)$, a candidate $c$}{Is $c$ a winner in $\calW$ according to rule $\calR$?}


\medskip

\noindent
The main results of this section are
the following: 

\begin{theorem}
\label{theorem:complexity-winner-determination-schulze}
The \textsc{Schulze Winner Determination} problem is $\nl$-complete.
\end{theorem}
\begin{theorem}
\label{theorem:complexity-winner-determination-rankedpairs}
The \textsc{Ranked Pairs Winner Determination} problem is $\ptime$-complete.
\end{theorem}

$\ptime$-membership in case of the ranked pairs method is obvious (and well-known). It remains to 
prove the remaining three properties: $\nl$-membership, $\nl$-hardness, and 
$\ptime$-hardness.

In the following hardness proofs, we will construct weighted tournament graphs $\calW=(A,E,\mu)$ with even integer weights and with $E$ being an asymmetric relation, i.e, if $(a,b)\in E$ then $(b,a)\notin E$. It follows from McGarvey's Theorem \citep{McGarvey} that such weighted tournament graphs can be obtained even from preference profiles containing only $\sum_{e\in E} \mu(e)$ many linear orders. 
Of course, our results also hold if preference profiles are given as input.

\medskip 

\noindent
Our {\bf NL-membership} proof 
for the Schulze-Winner Determination Problem
is based on the 
following two related problems $\textsc{WP}_{\geq}$ and  $\textsc{WP}_{>}$, where
we ask if, for given vertices $s,t$ and width $w$, a path from $s$ to $t$ of width $\geq w$
or $>w$, respectively, exists. Formally, we study the following problems:

\problem{Existence-of-Wide-Paths $\textsc{WP}_{\geq}$ / $\textsc{WP}_{>}$}{
a weighted graph $\calG$, vertices $s,t$, weight $w\in \mathbb{R}$}
{Does there exist a path from $s$ to $t$ \\
of width 
$\geq  w$ 
(in case of the $\textsc{WP}_{\geq}$-problem) or \\
of
width $> w$ (in 
case of the $\textsc{WP}_{>}$-problem)?}

%

\medskip

\noindent
It is straightforward to verify that both problems $\textsc{WP}_{\geq}$ and $\textsc{WP}_{>}$ are in $\nl$:
guess one vertex after the other of a path from $s$ to $t$ 
and check for any two successive vertices that they are connected by an edge of weight 
$\geq w$ or $> w$, respectively. 

Since $\nl = \conl$ by the 
famous Immerman-Szelepcs\'enyi Theorem, 
the co-problems of $\textsc{WP}_{\geq}$ and $\textsc{WP}_{>}$
are also in $\nl$. We can thus construct a non-deterministic Turing machine (NTM)
for the Schulze winner determination problem, which loops trough 
all candidates $c'\neq c$ and does the following:
\begin{itemize}[leftmargin=*]
\addtolength{\itemsep}{-2pt}
\item guess the width $w$ of the widest path from $c$ to $c'$ (among the weights of the edges in $\calW$);
\item solve the $\textsc{WP}{_\geq}$ problem for vertices $c,c'$ and weight $w$: 
check that 
there exists a path from $c$ to $c'$ of width $\geq w$;
\item  solve the co-problem of $\textsc{WP}_{>}$ for 
$c',c$ and weight $w$: 
check that 
there is {\em no} path from $c'$ to $c$ of width $> w$;
\end{itemize}
The correctness of the NTM is immediate. Moreover, 
by the above considerations on the problems 
$\textsc{WP}_{\geq}$,  $\textsc{WP}_{>}$ and their co-problems, 
the NTM clearly works in log-space time. 

Note that if we want to check if $c$ is the {\em unique\/} Schulze winner, 
then we just need to replace the third step above by a check that there is no path 
from $c'$ to $c$ of weight $\geq w$; in other words, we solve the co-problem
of $\textsc{WP}_{\geq}$. Again, the overall NTM clearly works in log-space.

\medskip 

\noindent
We prove {\bf NL-hardness}
by reduction from the 
$\nl$-complete 
Reachability problem\footnote{\citet{brandt09} show that any rule that selects winners from the Schwartz set---as Schulze's method does---and that break ties among candidates according to a fixed order is NL-hard to compute. Our result is similar but shows NL-hardness without requiring a tie-breaking order.}.
Let 
$(\calG,a,b)$ 
be an arbitrary instance 
of Reachability
with $\calG = (V,E)$ and $a,b \in V$.
We construct a weighted tournament graph $\calW = (V',E',\mu)$ 
from $\calG$ as follows and choose $a$ as the distinguished candidate: 

\begin{itemize}[leftmargin=*]
\addtolength{\itemsep}{-2pt}
\item First, remove from $\calG$ all edges of the form $(v, a)$ and $(b, v)$ for every $v \in V$, i.e., all incoming edges of 
$a$ and all outgoing edges from $b$.
\item For every pair of symmetric edges $e_1 = (v_i,v_j)$ and 
$e_2 = (v_j,v_i)$, choose one of these edges (say $e_1$)
and introduce a 
``midpoint'', i.e., add vertex $w_{ij}$ to $V$ and replace 
$e_1$ by the two edges $(v_i,w_{ij})$ and $(w_{ij},v_j)$.
\item For every vertex $u$ different from $b$, introduce a 
new vertex $r_u$ and edges $(b, r_u)$, $(r_u, u)$. 
\item Now there is exactly one incoming edge of $a$, 
namely $e = (r_a,a)$. 
Define the weight of this edge as $\mu(e) = 2$ and set 
$\mu(e') = 4$
for every other edge $e' \neq e$.
\end{itemize}
It is easy to verify that $a$ is a Schulze winner in $\calW$ (actually, 
it is even the unique Schulze winner), if and only if there is a path from $a$ to $b$ in $\calG$. 
To see this, first observe that there is a path from $a$ to $b$ in $\calW$ if and only if there is one in $\calG$. Hence, if there is a path from $a$ to $b$, then the widest path from $a$ to any vertex in $\calW$ is $4$. Conversely, all paths from any vertex to $a$ must go through edge $(r_a,a)$ and, therefore, have width at most $2$. 
On the other hand, if there is no path from $a$ to $b$, then $a$ cannot be a winner, since $b$ indeed has a path to $a$ (via $r_a$) and, 
therefore, in this case, 
$b$ is definitely preferred to candidate $a$ according to the Schulze method.

\medskip 

\noindent
The {\bf P-hardness}
proof for the Ranked Pair Winner Determination problem\footnote{The proof in the original version of this paper \citep{ijcai/CsarLP-schulze} is incorrect. This error was identified and communicated to us by Zack Fitzsimmons, Zohair Raza Hassan, and Edith Hemaspaandra, who also provided the corrected proof in this technical report, included here with their kind permission.} relies on the Edge Maximal Acyclic Subgraph (EMAS) problem \citep{greenlaw1992parallel,greenlaw1995limits}:

\problem{Edge Maximal Acyclic Subgraph (EMAS)}{
	a directed graph $\calG=(V,E)$ with an ordering over edges, a designated edge~$f$}
{Is $f$ contained in the edge maximal acyclic subgraph (EMAS)?
That is, edges are added iteratively to a set in the given order, omitting those that would introduce a cycle; is $f$ being added to this set?}

We show P-hardness by reducing the complement of EMAS (accept if the given edge $f$ is not contained in the EMAS) to our problem. Note that it is fine to reduce from the complement since $\coP = \Ptime$.
Let $\calG=(V,E)$ be a given digraph with $E=\{e_1,\dots,e_m\}$, and edges are ordered $e_1$, $e_2$, etc. Without loss of generality, we assume that $\calG$ has no self-loops $(v,v)$ and no parallel edges $(u,v),(v,u)$, as these could easily be handled in a preprocessing step in logspace  by subdividing those edges while preserving lexicographic order.

We construct a weighted tournament $\calW=(A,F,\mu)$ as follows.
\begin{enumerate}
	\item Let $A=V\cup\{0\}$. 
	\item For each edge $e_i\in E$ with $e_i\neq f$, we add $e_i$ to $F$ with weight $\mu(e_i)=m-i+1$.
	\item Suppose $f=(u,v)$ with label~$j$ (i.e., $f=e_j$). We add edges $(u, 0)$ with weight $\mu((u, 0)) = m - j + 1$ and $(0, v)$ with weight $m + 1$.
	\item We assign weight 0 to all other edges.
\end{enumerate}
We assume lexicographic tiebreaking for selecting edges.
In particular, edge $(0, w)$ is picked before $(w, 0)$ for all $w\neq u$.
This reduction is clearly possible in logspace.

We claim that vertex $0$ is a winner according to ranked pairs if and only if $f$ is not contained in the EMAS of $\calG$.

First, note that when evaluating this tournament with the ranked pairs method, edge $(0, v)$ is picked first and then
the edges from $E$ are picked in the same order as in the EMAS problem. Moreover, $(u, 0)$ is picked when $f$ would have been picked. Note that cycles in $\calG$ correspond to cycles in $\calW$ and vice versa.

To show the first direction of the claim, assume that $0$ is the winning candidate. Then $(u, 0)$ was not picked due to a cycle in $\calW$. Consequently, $f$ is not contained in the EMAS, as there would be a cycle involving $f$ in $\calG$.

Conversely, if $f$ is not contained in the EMAS, there is a cycle involving $f$. This translates to a cycle involving $(u, 0)$ and $(0, v)$ in $\calW$, and hence $(u, 0)$ will not be picked. Consequently, $(0, u)$ is picked as well as $(0, w)$ for all $w\neq u$ due to tiebreaking. We see that $0$ is the winning candidate.

\section{Computation of the Schulze Winner}
\label{sect:schulze}
\def\ws{\mathit{ws}}
\def\wt{\mathit{wt}}


In this section, we present our new Pregel-based algorithm for determining the 
Schulze winners for a given weighted tournament graph $\calW = (A, E, \mu)$.
A straightforward algorithm,
which is implicit in the $\nl$-membership proof in Section~\ref{sect:complexity}, 
would consist in computing for every pair $(a,b)$ of vertices in $\calW$
the widths $p(a,b)$ and $p(b,a)$.
However, this would mean that we have to store for each vertex $c$ a linear amount of information (namely $p(c,v)$ for every $v\in A$). This 
contradicts the philosophy of Pregel algorithms which 
aim at keeping the local information at each vertex small \citep{yan2014pregel}.

An improvement of this idea would be to first compute the SCCs of $\calW$
by one of the Pregel algorithms in the literature 
\citep{salihoglu2014optimizing,yan2014pregel} and to compute the 
widest paths for pairs of vertices only for the non-dominated SCCs (i.e., the Schwartz winners). This approach has the disadvantage that the set of pairs 
to be considered may still be big and, moreover, the computation of the SCCs only makes use of part of the information (namely the dominance graph---thus ignoring the weights of the edges). 

We therefore construct a new Pregel-style algorithm, which draws some inspiration from the well-studied SCC computation (in particular, the forward/backward propagation of minimum vertex-ids) but utilises the weight information to prune the search space as soon as possible. 

The proposed Pregel algorithm for computing the Schulze Method is guaranteed to have very small local information at each vertex. The overall structure of our algorithm is given in 
Algorithm~\ref{alg:schulze-overall}.

\begin{algorithm}
\caption{Schulze Winner Determination}
\label{alg:schulze-overall}
	\begin{algorithmic}
  	\small
  	\vspace{2pt}
  	\State{Initialisation-of-vertices};
        \While{there exists a vertex $c$ with $c.\status$ = `unknown'}
            \State{Preprocessing}; 
            \State{Forward-Backward-Propagation};
            \State{Postprocessing}; 
	\EndWhile
  	\State{Output vertices with status `winner'};
	\end{algorithmic}
\end{algorithm}

We assume that each vertex $c$ 
is assigned a unique id $c.\id \in \{1, \ldots m\}$.
Moreover, each vertex $c$ has a $\status$ which may take one of 3 values
$\{$`winner', `loser', `unknown'$\}$ to express that 
$c$ is a Schulze winner, not a Schulze winner, or if we do not know yet, respectively. 
Additional information stored at each vertex includes the
fields $s,t,\ws, \wt$, and $\scc$, whose meaning will be explained below, as 
well as information on the adjacent edges together with their weights.

In {\bf Initialisation-of-vertices},
we determine for each candidate $c$ 
the maximum weight of all incoming and outgoing edges and set the status accordingly: 
if \[\max_{a \in A} \mu(a,c) > \max_{a \in A} \mu(c,a),\] then 
there exists a vertex $v$ (namely the one with 
$\mu(v,c) = \max_{a \in A} \mu(a,c)$) which is preferred to $c$ by the Schulze method. 
Hence, in this case, we  set $c.\status$ = `loser'; 
otherwise we set $c.\status$ = `unknown';

The goal of each iteration of the {\bf while-loop} in Algorithm~\ref{alg:schulze-overall} is 
to compute for every vertex $c$ the ids 
$s$ (= source) and
$t$ (= target) which are the minimum ids among all vertices with $\status$ = `unknown'
such that there is a path from $s$ to $c$ and from $c$ to $t$. Moreover, we
also determine the weights $\ws$ and $\wt$ of the widest paths from  
$s$ to $c$ and from $c$ to $t$.
Termination is guaranteed since in each iteration 
at least one vertex 
changes its status from `unknown' to 
`loser' or `winner'.
Our experimental evaluation  
shows that the algorithm terminates very fast: on real-world data, 
typically even a single iteration of the while-loop suffices. 
Preliminary experiments with synthetic data show that 
the while-loop is executed less than 10 times 
for instances with 10.000 candidates.

In {\bf Preprocessing},
 shown in Algorithm \ref{alg:preprocessing}, 
we initialise the fields $(s,\ws,t,\wt)$ of all vertices. 
For every vertex $c$ 
with $c.\status$ = `unknown', we set $c.s = c.t = c.\id$. 
Thus, initially,
the minimum id of vertices to reach $c$ and reachable from $c$ is 
the id of $c$ itself. 
We send the information on 
$c$ as a source (resp.\ target) to its adjacent vertices via outgoing (resp.\ incoming) edges. 
For vertices with status different 
from `unknown', we set $c.s = c.t = \infty$. 
This allows such a vertex $c$ to pass on vertex ids 
from other sources and targets but it prevents $c$ from 
passing on its own id.

\begin{algorithm}
\caption{Preprocessing}
\label{alg:preprocessing}
	\begin{algorithmic}
  	\small
  	\vspace{4pt}
  	\If{$c.\status$ = `unknown'}
            \State{$s$ = $c.id$; $\ws$ = $\infty$; 
             $t$ = $c.id$; $\wt$ = $\infty$}; 
            \State{\mbox{}\hskip0pt{\bf for each} 
               outgoing edge $(c,v)$ with weight $w$ {\bf do}} 
            \State{\mbox{}\hskip12pt send $($`forward'$,s,w)$ to vertex $v$}; 
            \State{\mbox{}\hskip0pt{\bf for each} 
               incoming edge $(v,c)$ with weight $w$ {\bf do}} 
            \State{\mbox{}\hskip12pt send $($`backward'$,t,w)$ to vertex $v$}; 
        \Else
            \State{$s$ = $\infty$; $\ws$ = $0$; 
            $t$ = $\infty$; $\wt$ = $0$};         
	\EndIf
	\end{algorithmic}
\end{algorithm}

\begin{algorithm}
\caption{Forward-Backward-Propagation}
\label{alg:propagate}
	\begin{algorithmic}
  	\small
  	\vspace{4pt}

  	\State{\mbox{}\hskip0pt{\bf for each} 
                    received value $(d,v,w)$ {\bf do}}; 
            \State{\mbox{}\hskip12pt {\bf if} $d = $ `forward' {\bf then}}                   
                \State{\mbox{}\hskip24pt {\bf if} $v < s$ {\bf then}
                    $s = v$; $\ws = w$};                
                \State{\mbox{}\hskip24pt {\bf else if} $v = s$ {\bf then}
                    $\ws$ = $\max(\ws,w)$}; 

            \State{\mbox{}\hskip12pt {\bf else if} $d = $ `backward' {\bf then}}                   
                \State{\mbox{}\hskip24pt {\bf if} $v < t$ {\bf then}
                    $t = v$; $\wt = w$};                
                \State{\mbox{}\hskip24pt {\bf else if} $v = t$ {\bf then}
                    $\wt$ = $\max(\wt,w)$}; 
                                 
                
            
\vspace{3pt}
                \State{\mbox{}\hskip0pt 
                   {\bf if} $(s,\ws)$ has changed {\bf then}}            
                     \State{\mbox{}\hskip12pt{\bf for each} 
               outgoing edge $(c,v)$ with weight $w$ {\bf do}} 
            \State{\mbox{}\hskip24pt send 
               $($`forward'$,s,\min(\ws,w))$ to vertex $v$}; 
\vspace{3pt}            
                \State{\mbox{}\hskip0pt 
                   {\bf if} $(t,\wt)$ has changed {\bf then}}            
                     \State{\mbox{}\hskip12pt{\bf for each} 
               incoming edge $(v,c)$ with weight $w$ {\bf do}} 
            \State{\mbox{}\hskip24pt 
            send $($`backward'$,t,\min(\wt,w))$ to vertex $v$}; 
\vspace{3pt}            
            \State{set $c$ inactive}; 
	\end{algorithmic}
\end{algorithm}

The {\bf Forward-Backward-Propagation} is the actual 'Pregel heart' of the computation. 
The other procedures work in parallel too, but they do not use the Pregel Computation API.
Algorithm \ref{alg:propagate} realizes the forward and backward  propagation as a Pregel procedure.  
For each vertex $c$, 
we determine (1) the minimum source-id $s$ together with the 
maximum width $\ws$ of paths from $s$ to $c$ and (2) 
the minimum target-id $t$ together with the 
maximum width $\wt$ of paths from $c$ to $t$.
We thus analyse each received message $(d,v,w)$ consisting of a direction $d$,
vertex id $v$, and width $w$. 
In case of `forward' direction, 
we have to check if we have found a source $v$ with a yet smaller id
than the current value $s$. If so, we update $c.s$ and $c.\ws$ accordingly. If the received vertex-id $v$ is equal to the current value of $c.s$, we have to update 
$c.\ws$ in case the received value $w$ is greater than $c.\ws$ 
(i.e., from the same source we have found a path of greater width).

Messages in `backward' direction are processed analogously, resulting in 
possible updates of the target-id $t$ and/or the width $\wt$ of paths from $c$ to $t$. 
After all messages have been processed, we propagate the 
information on new source id $s$ (resp.\ target id $t$) and/or increased width
of paths from $s$ to $c$ (resp.\ from $c$ to $t$) to all adjacent vertices of $c$ in forward (resp.\ backward) direction. 
Forward-Backward-Propagation terminates when no more messages 
are pending.

In {\bf Postprocessing}, 
as shown in Algorithm \ref{alg:postprocessing},
we use two crucial properties of  source and target ids, 
which are inherited from the SCC computation 
in \citep{yan2014pregel}: 
First, if for a vertex $c$, we have $c.s = c.t$, then the set of vertices $v$ with the same source/target id (i.e., 
$v.s = v.t = c.s$) forms the SCC of $c$. 
Second, if for two vertices $c$ and $v$, we have $c.s \neq v.s$ or $c.t \neq v.t$, then 
$c$ and $v$ belong to two different SCCs.

\begin{algorithm}
\caption{Postprocessing for vertex $c$}
\label{alg:postprocessing}
	\begin{algorithmic}
  	\small
  	\vspace{2pt}

  	\If{$c.s < c.t$}
  	   \State{$c.\status$ = `loser'};
  	\ElsIf{$c.s > c.t$}
  	   \State{set $\status$ of vertex $c.t$ to `loser'};
  	\ElsIf{$c.s = c.t$}
  	   \State{$c.\scc = c.s$}; 
       \State{\mbox{}\hskip0pt 
                   {\bf if} $c.\ws > c.\wt$ {\bf then} $c.\status$ = `loser'};            
       \State{\mbox{}\hskip0pt 
                   {\bf else if} $c.\ws < c.\wt$ {\bf then} 
                   set $\status$ of vertex $c.s$ to `loser'};              	
    \EndIf
  	\vspace{2pt}

  	\For{{\bf each} incoming edge $(v,c)$}
  	   \State{get $(v.s,v.t)$ from vertex $v$};
  	   \If{$c.s \neq v.s$ \textbf{or} $c.t \neq v.t$}
  	      \If{$c.s = c.t$}
  	      \State{{\bf for each} vertex $u$ with $u.\scc = c.s$ {\bf do}}
            \State{\mbox{}\hskip12pt 
  	           set $\status$ of vertex $u$ to `loser'};
  	      \ElsIf {$c.s \neq c.t$} 
  	        \State{$c.\status$ = `loser'};
  	      \EndIf
  	   \EndIf 	   
  	\EndFor  	 	
    \vspace{2pt}
  	   
    \If{$c.\scc = c$ {\bf and} $c.\status$ = 'unknown'}
  	   \State{$c.\status$ = `winner'};
    \EndIf 	   
  	
	\end{algorithmic}
\end{algorithm}

In Algorithm \ref{alg:postprocessing}, 
we first compare for each vertex $c$ the values of $s$ and $t$: 
if $c.s < c.t$, then  $c$ is reachable from $s$ but $s$ is not reachable from $c$. Hence, $c$ is a loser. If 
$c.s > c.t$, then  $t$ is reachable from $c$ but $c$ is not reachable from $t$. Hence, $t$ is a loser. Note that setting the status of $t$ (which is, in general, different from the current node) is done by a subroutine whose details are omitted here. 
Finally, if $c.s = c.t$, then 
(as recalled above) we have found the SCC of $c$. As in \citep{yan2014pregel}, 
we use the minimum id of the vertices in an SCC to label the SCC. 
If the width of the path from $s$ (which is equal to $t$ by our case distinction) to $c$ is greater than from $c$ to $t$, then $c$ is a loser (since $s$ is preferred to it). In the opposite case, $s$ is~a~loser. 

In the next step in Algorithm \ref{alg:postprocessing}, we compare the values of $(s,t)$ of each vertex $c$ with 
the values of $(v.s,v.t)$ of all vertices $v$ with an edge $(v,c)$. If 
$v.s \neq c.s$ or $v.t \neq c.t$, then 
$v$ and $c$ are in different SCCs. By the existence of the edge $(v,c)$, this means that there can be no path from $c$ to $v$. 
Hence, $v$ is preferred to $c$ according to the Schulze method. Moreover, if 
$c.s = c.t$, then we have found the SCC of $c$. In this case, $v$ is preferred to all vertices $u$ in this SCC. 

Finally, suppose that we have found some SCC such that the vertex $c$ with minimum id in this SCC has not been identified as a loser by any of the above cases. In particular, this means that none of the vertices in this SCC has an incoming edge from outside the SCC and, 
moreover, the SCC cannot contain a vertex $v$ with $p(v,c) > p(c,v)$. 
In this case, we may mark vertex $c$ 
as a winner. It is now also clear that at least one vertex must change its status from `unknown' to either 'loser' or `winner' 
in every execution of Algorithm \ref{alg:postprocessing} and, therefore, in every iteration of the while-loop of 
Algorithm \ref{alg:schulze-overall}.

We briefly describe further {\bf optimisations}.
For instance, rather than assigning vertex-ids randomly, 
we first 
perform the cheap computation of Borda scores and assign vertex-ids in 
descending order of Borda scores.
This yields lower ids for vertices that are more likely to dominate other vertices, thus speeding up the exclusion of dominated candidates.
Moreover, in the postprocessing phase, 
we exclude all vertices of a dominated SCC from further consideration simply by setting the weights of all incoming edges to this SCC to 0.
This is much faster than physically deleting these vertices, since  in GraphX, changing 
the graph topology is time intensive.

Finally, we remark that by iterating the algorithm $k$ times and continuously removing Schulze winners, it is straight-forward to compute a top-$k$ ranking according to Schulze.

\section{Experiments}
\label{sect:experiments}
\subsection{Setup and Data}

To assess the performance of big data algorithms, it is essential to test them against actual real-world large-scale data sets.
To this end, 
we use the Spotify ranking data\footnote{\url{https://spotifycharts.com/regional}} of 2017, which consists of daily top-200 music rankings for 53 countries.
We 
consider the ranking of each day and country as a single voter, and 
then generate the corresponding weighted tournament graph. 
Our experiments are based on four data sets generated from this data: Global150, Global200, Europe150, and Europe200, which are based on daily top-150/top-200 charts of all available/European countries.
We do not take into account the number of listeners in each country, since this information is not available---but this could easily be included in our computation by assigning countries weights that correspond to the number of listeners. 
In Table~\ref{tab:data}, we provide an overview of these four data sets.
All four weighted tournament graphs are dense: edges exist between roughly 94\% of all candidate pairs.
We note that the Spotify data sets used here are significantly larger than any instances available in the PrefLib database~\citep{mattei2013preflib} and the data sets used by \citet{AAAI17}, which have $\leq 7000$ vertices and less than 5 million edges.

{
\setlength{\tabcolsep}{4pt}
\begin{table}
\caption{Spotify data sets}
\centering
\begin{tabular}{rccccccccccc}
\toprule
  & {candidates} & {voters}  & {edges} & {after preproc.}\\\midrule
Europe150   & 9698 &  7,481 & 44.1M & 11 undecided\\
Europe200  &  12250   & 7,481 & 70.7M& 12 undecided \\
Global150   & 14187 & 15,553   & 94.9M& 8 undecided\\
Global200 &   18407 & 15,553   & 159.6M& 9 undecided \\ 
\bottomrule
\end{tabular}
\label{tab:data}
\end{table}
}


We ran our experiments on a Hadoop cluster with 18 nodes (each with an Intel Gold 5118 CPU, 12 cores, 2,3 GHz processor, 256 GB RAM, and a 10Gb/s network connection).
To better observe the scalability of our algorithm, we restricted the number of cores and nodes (details follow).

Our Schulze algorithm is implemented in the Scala programming language.
Furthermore, we use the GraphX library\footnote{\url{https://spark.apache.org/graphx/}}, which is built on top of Spark
\citep{zaharia2010spark}, an open-source cluster-computing engine.
GraphX provides a Pregel API, but is slightly more restrictive than the Pregel framework.
In particular, in GraphX, only messages to adjacent vertices can be sent, while 
other Pregel implementations allow messages to be sent to arbitrary vertices.
The source code of our implementation is part of the open-source project CloudVoting\footnote{\url{https://github.com/theresacsar/CloudVoting}}.

\subsection{Results}
\begin{figure}
\includegraphics[width=.9\columnwidth]{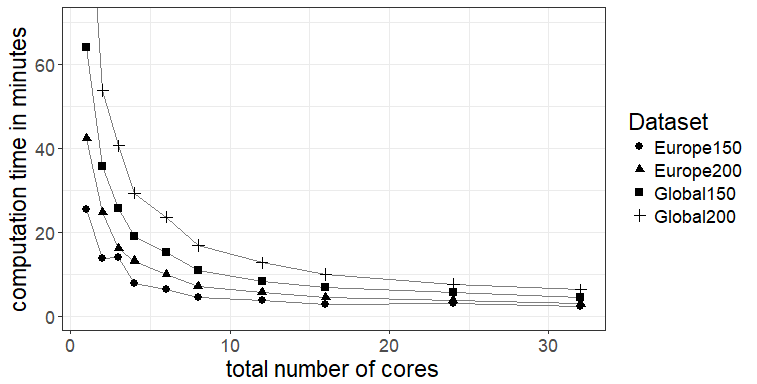}
\caption{Runtime required for computing Schulze winners.}
\label{fig:runtime}
\end{figure}

Our experiments show that our algorithm scales very well with additional computational resources: both an increase in nodes and in cores per node significantly sped up the computation.
We refer the reader to Figure~\ref{fig:runtime} for an overview of runtimes for 1/2/3/4 nodes with 1/2/4/8 cores each.
On the x-axis of this chart we show the total number of cores, i.e., the number of nodes times the number of cores per node. 
For x-values with multiple interpretations we show the best runtime.
This is always the configuration with most cores per node, but the differences between an increase in nodes or cores is almost negligible.
Furthermore, our implementation manages to compute Schulze winners of all Spotify data sets within very reasonable time: with 4 nodes each using 8 cores,
the data sets could be handled in less than 6.5min. 

It is insightful to compare the runtimes of our algorithm with the original, sequential algorithm~\citep{Schulze} based on the classical Floyd--Warshall algorithm; to the best of our knowledge this is the only published algorithm computing the Schulze method.
These two algorithms differ not only in their capability of parallelization, but also in that our algorithm only returns Schulze winners whereas the original algorithm returns a full ranking of candidates.
Due to our focus on Schulze winners, we could include the many optimisations described in Section~\ref{sect:schulze}.
As a consequence of these optimisations and our focus on winners, our algorithm is faster than the original algorithm even without parallelization (1 node with 1 core):
Our algorithm requires with this configuration 26min/42min/64min/105min for the the Europe150/200 and Global150/200 data sets, respectively.
In contrast, the original algorithm (also implemented in Scala) requires
71min/143min/221min for the Europe150/200 and Global150 data sets; it did not terminate in reasonable time for the Global200 data set and we stopped the computation after 5 hours.
As mentioned before, this comparison is not completely fair due to the different output, but shows the impact of the optimisations in our algorithm.

We also performed preliminary experiments with low-density graphs (based on synthetic data). 
We observed that for these graphs the number of undecided candidates decreases slower with each iteration of the forward-backward propagation, in contrast to the Spotify data sets where even after preprocessing very few undecided candidates remained (cf.~Table~\ref{tab:data}).
However, for 10,000 candidates we obtain comparable runtimes to the Spotify data sets and hence are optimistic that our algorithm behaves rather robustly with respect to varying densities.

\section{Conclusion and Directions for Future Work}
\label{sect:conclusion}

This paper shows the great potential of parallel algorithms and cloud computing techniques in computational social choice, but many challenges remain. During the experimental evaluation we experienced that it is a non-trivial task to generate synthetic data sets sufficiently large for benchmark purposes. In particular, it is challenging to generate preference profiles according to the widely used Mallows model with more than 10,000 candidates.
The compilation of large real-world data sets is equally important, as most preference data sets currently available are insufficient for large-scale experiments. 
As further future work we plan to investigate other voting rules with respect to their parallelizability and their suitability for handling large preference data sets. 


\section*{Acknowledgments}
This work was supported by the Austrian Science Fund projects (FWF):P25518-N23, (FWF):P30930-N35 and (FWF):Y698. Further we would like to thank G\"{u}nther Charwat for his feedback on the experimental section.

We are very grateful to Zack Fitzsimmons, Zohair Raza Hassan, and Edith Hemaspaandra, who identified the error in our original P-hardness proof for the ranked pairs method and generously provided us with a correct proof of the statement.

\bibliographystyle{abbrvnat}
\bibliography{bibliography}

\end{document}